\documentclass{elsart}
\usepackage[tbtags]{amsmath}
\usepackage{amsfonts,graphicx}
\journal{quant-ph/0603169}
\usepackage[latin1]{inputenc}

\newcommand{\ket}[1]{|#1\rangle}
\newcommand{\bra}[1]{\langle#1|}

\DeclareMathOperator{\tr}{Tr} 
\DeclareMathOperator{\id}{\mathbf{1}}

\SetSymbolFont{symbols}{normal}{OMS}{cmsy}{m}{n}

\begin{document}
\begin{frontmatter}
\boldmath
\title{An open quantum system approach to EPR~correlations 
       in $K^0 \bar{K}^0$ system}
\unboldmath

\author{Pawe{\l}{} Caban},
\ead{P.Caban@merlin.fic.uni.lodz.pl}
\author{Jakub Rembieli{\'n}ski},
\ead{J.Rembielinski@merlin.fic.uni.lodz.pl}
\author{Kordian A. Smoli{\'n}ski},
\ead{K.A.Smolinski@merlin.fic.uni.lodz.pl}
\author{Zbigniew Walczak},
\ead{Z.Walczak@merlin.fic.uni.lodz.pl}
\author{Marta W{\l}odarczyk}
\ead{M.Wlodarczyk@merlin.fic.uni.lodz.pl}
\address{Department of Theoretical Physics, University of Lodz\\
Pomorska 149/153, 90-236 {\L}{ó}d{\'z}, Poland}

\date{\today}
\begin{abstract}
We find the time evolution of the system of two non-interacting
unstable particles, distinguishable as well as identical ones, 
in arbitrary reference frame having only the Kraus operators 
governing the evolution of its components in the rest frame.
We than calculate in the rigorous way Einstein--Podolsky--Rosen quantum 
correlation functions for $K^0 \bar{K}^0$ system in the singlet 
state taking into account $CP$-violation and decoherence and show 
that the results are exactly the same despite the fact we treat kaons 
as distinguishable or identical particles which means that the
statistics of the particles plays no role, at least in considered cases.
\end{abstract}
\maketitle
\end{frontmatter}

\section{Introduction}
In the recent years the possibility of testing Bell-CHSH inequalities 
\cite{bell64,chsh} in the system of correlated neutral mesons 
has attracted some attention 
(see e.g. \cite{bertlmann06,bramon05,genovese05,go04,bertlmann04,%
bramon02a,dalitz01,gisin01,bertlmann01,bertlmann01a,hiesmayr01,%
bramon99,ancochea99,benatti98a,apostolakis98:abb,uchiyama97,%
domenico95}), because kaons and $B$-mesons are detected with higher 
efficiency than photons which maybe allows one to close the detection
loophole (see e.g. \cite{genovese05}). 

The crucial point in studying quantum correlations in the system of
unstable particles, say $K^0 \bar{K}^0$, is the choice of a model
describing time evolution of the system under consideration. 
It is obvious that this choice depends on what physical aspects 
we want to neglect and what we want to focus on. For example, if we  
want to focus our attention on decay of particles governed by the
Geiger--Nutall law neglecting the evolution of decay products and other
physical processes like e.g. decoherence, we can use the standard
Weisskopf--Wigner approach \cite{weisskopf30a,weisskopf30b}. 
The price we must pay for this choice are some ambiguities, 
due to the non-Hermitian Hamiltonian of Weisskopf--Wigner model,
when we want to consider a series of consecutive measurements.
However, if we mainly want to take into accout decoherence
we can do our calculations in the framework of many models considered 
in the literature 
(see e.g. \cite{bertlmann03,andrianov01,benatti97a,ellis96}). 
The common feature of all these models is taking into account only the
two-particle sector of density matrix describing the state of the
$K^0 \bar{K}^0$ system, neglecting the one-particle and zero-particle
sectors which arise during the time evolution due to the decay process.

Recently, we have introduced a model \cite{caban05} of time evolution
of a neutral kaon in its rest frame, which allows one to describe 
the $K^0 \bar{K}^0$ system in more realistic way. 
In the framework of our model, based on the theory of open quantum
systems (see e.g. \cite{breuer02,alicki01}),
we can take into account that the state of the system evolves from 
a two-particle state to the zero-particle one through states being 
a mixture of all possible states. 
This model is based on the following assumptions.
(i) A neutral kaon is treated as an open quantum system, 
and the decay products are regarded as a part of the environment. 
(ii) The system can be found either in a particle
state, or in the state of absence of the particle, which we call 
(in analogy to the quantum field theory) 
the vacuum state, and denote by $\ket{0}$.  
Therefore the Hilbert space of the neutral kaon--vacuum system 
is a direct sum of one-particle Hilbert space and one-dimensional 
space $\mathcal{H}_0$ spanned by $\ket{0}$. 
(iii) The superselection rule prohibiting the superposition of the
particle and vacuum holds. 
(iv) The time evolution of the system, consistent with the
Geiger--Nutall law and allowing for $CP$ symmetry violation, 
is given by a family of completely positive trace 
preserving maps forming a one-parameter dynamical semigroup. 
Complete positivity implies that the time evolution of a state of the
system can be written in the operator-sum (Kraus) representation 
\cite{kraus} (which immediately allows one to find the time
evolution of non-interacting particles)
\begin{equation}
\hat{\varrho}(t)=\sum_{i}\hat{E}_{i}(t)\hat{\varrho}(0)\hat{E}^{\dag}_{i}(t),
\label{kraus}
\end{equation}
and the trace preservation requirement leads to the condition
\begin{equation}
\label{normalizacja}
\sum_{i}\hat{E}^{\dag}_{i}(t)\hat{E}_{i}(t)=\id.
\end{equation} 
As a remarkable side effect of complete positivity of time evolution 
of the system we obtained the upper bound for the decoherence
parameter (see \cite{caban05}).
Here we would like to point out that we used the dynamical semigroup
approach for the entire evolution of the system, not only for the
description of its decoherence, as was done in 
\cite{bertlmann03,andrianov01,benatti97a,ellis96}.

The paper is organized as follows. In Section \ref{ruchoma czastka} we
generalize our description of time evolution of an unstable particle 
in its rest frame to the case of an arbitrary reference frame. 
Then, in Section \ref{two unstable particles}, we show how the
evolution of the system of two unstable non-interacting particles 
can be found when we know the Kraus operators governing the
evolution of its components, both for distinguishable  and
identical particles case. 
Finally, in Section \ref{correlations} we apply these results in
calculation of some quantum correlation functions for 
$K^0 \bar{K}^0$ system in the singlet state. 
And we show that the obtained results are exactly the same for kaons
treated as distinguishable or identical particles.

\section{Time evolution of an unstable particle in an arbitrary
reference frame}
\label{ruchoma czastka} Let us assume that we know the Kraus
representation of time evolution of 
an unstable particle in its rest frame 
in which the four-momentum is $\tilde{k}=(m,\vec{0})$. 
Let us answer
the question, how this evolution looks like from the point of view
of an observer in the laboratory frame, in which the particle has
four-momentum $k=(k^0,\vec{k})$. Let $L_k$ be the Lorentz boost
from the rest frame to the laboratory frame, i.e.~$k=L_k \tilde{k}$.
When we denote the inner degrees of freedom (e.g.~strangeness,
bottom) as $s$, than the state of particle with four-momentum $k$
can be denoted as $\ket{s,k}$ (in this notation,
$\ket{0}\equiv\ket{0,0}$). Since we deal only with (pseudo)~scalar
particles (e.g.~kaons), $\ket{s,k}=U(L_k)\ket{s,\tilde{k}}$, which
implies that the density operator in laboratory frame has
the following form
\begin{equation}
\begin{split}
\hat{\varrho}^{{\rm
lab}}(t)&=U(L_k)\hat{\varrho}(\tau^k)U^{\dag}(L_k)\\ &=
 \sum_i\hat{E}^{k\,{\rm lab}}_i(\tau^k)\hat{\varrho}^{{\rm lab}}(0)
(\hat{E}^{k\, {\rm lab}}_i(\tau^k))^{\dag},
\end{split}
\label{ewolucjaa}
\end{equation}
where the proper time in the rest frame of the particle is denoted by
$\tau^k$ because it is convenient to express it by means of $k$ and the
laboratory time $t$:
\begin{equation}
\tau^k=\frac{t}{\sqrt{1+\frac{\vec{k}^2}{m^2}}}\equiv\frac{t}{\gamma^k},
\label{proper time}
\end{equation}
and $\hat{E}^{k\,{\rm lab}}_i(\tau^k)=U(L_k) \hat{E}_i(\tau^k)
U^{\dag}(L_k)$. By means of the decomposition
$\hat{E}_i(\tau^k) = \sum_{s,s'}
E_i^{ss'}(\tau^k)\ket{s,\tilde{k}}\bra{s',\tilde{k}}$ (see
Appendix~\ref{AppA}), we get
\begin{equation}
\hat{E}^{k\,{\rm lab}}_i(\tau^k)=
\sum_{s,s'}E_i^{ss'}(\tau^k)\ket{s,k}\bra{s',k}.
\end{equation}
It means that the matrix elements of $\hat{E}^{k\,{\rm
lab}}_i(\tau^k)$ in the basis $\{\ket{s,k}\}$ are exactly the same
as the matrix elements of $\hat{E}_i(\tau^k)$ in the basis
$\{\ket{s,\tilde{k}}\}$. Note that this holds only for (pseudo)
scalar particles.

Taking into account the form of the Kraus operators given in
Appendix~\ref{AppA}, the evolution (\ref{ewolucjaa}) can be easily
extended to the case in which the particle state is the
superposition (or mixture) of a few different momentum eigenstates.
In this case the space of states of the system is spanned by the
orthogonal vectors $\ket{s,k}$ and vacuum, but now the four-momentum
$k$ can take an arbitrary but finite number of different values. We
denote the set of admissible four-momenta by $Q$ (for example, in
the case of two identical particles discussed in the next section,
$Q$ consists of only two elements $Q=\{p,q\}$).  The time evolution
of the density operator describing such a state has the following
form
 \begin{equation}
   \begin{split}
 \hat{\varrho}^{{\rm lab}}(t) &= \hat{E}_{0}\hat{\varrho}^{{\rm lab}}(0)
 \hat{E}_{0}^\dag\\ &\quad
  + \sum_{k\in Q}\sum_{i=1}^{5}\hat{E}^{k\,{\rm lab}}_{i}(\tau^k)
 \hat{\varrho}^{{\rm lab}}(0)
 (\hat{E}^{k\, {\rm lab}}_i(\tau^k))^\dag,
\end{split}
\label{ewolucja_wiele_pedow}
 \end{equation}
where the Kraus operators  $\hat{E}_{0}=\hat{E}_{0}^{\rm
lab}=\ket{0,0}\bra{0,0}$ and $\hat{E}^{k\,{\rm lab}}_{i}(\tau^k)$,
 are given in Appendix \ref{AppA}.
 Note that when $Q$ consists of only one element
Eq.~(\ref{ewolucja_wiele_pedow}) reduces to (\ref{ewolucjaa}).

\section{Two unstable particles as open quantum system}
\label{two unstable particles} Let us consider two
unstable particles $\mathcal{A}$ and $\mathcal{B}$ (this approach can
be easily extended to the multiparticle case), with four-momenta $p$ and
$q$ in the laboratory frame, respectively. The space of states of the
whole system is
\begin{equation}
\label{hilbert}
\begin{split}
\mathcal{H}&=\left(\mathcal{H}_{\mathcal{A}}\oplus\mathcal{H}_{0}\right)
\otimes
\left(\mathcal{H}_{\mathcal{B}}\oplus\mathcal{H}_{0}\right)\\ &=
(\mathcal{H}_{\mathcal{\mathcal{A}}}\otimes\mathcal{H}_{\mathcal{B}})
\oplus(\mathcal{H}_\mathcal{A}\otimes\mathcal{H}_0
\oplus\mathcal{H}_0\otimes\mathcal{H}_\mathcal{B})\oplus(\mathcal{H}_0
\otimes\mathcal{H}_0),
\end{split}
\end{equation}
so one can easily see that the system can be either in two-particle
state ($\mathcal{H}_{\mathcal{A}}\otimes\mathcal{H}_{\mathcal{B}}$),
or in one-particle state
($\mathcal{H}_\mathcal{A}\otimes\mathcal{H}_0\oplus\mathcal{H}_0
\otimes\mathcal{H}_\mathcal{B}$),
or in the vacuum state $(\mathcal{H}_0\otimes\mathcal{H}_0)$. Such a
system is an open one, so its evolution is represented by a
completely positive map and can be written in the operator-sum
representation
\begin{equation}
\label{krausABa} \hat{\varrho}_{\mathcal{A}\mathcal{B}}^{{\rm
lab}}(t)= \sum_{k\in Q,k'\in Q'}\sum_{i,j}\hat{E}_{ij}^{kk'}(t)
\hat{\varrho}_{\mathcal{A}\mathcal{B}}^{{\rm
lab}}(0)(\hat{E}_{ij}^{kk'}(t))^{\dag},
\end{equation}
where, under a reasonable assumption that the particles do not
interact, we define
\begin{equation}
\label{krausy} \hat{E}_{ij}^{kk'}(t)=\hat{E}^{k\,{\rm
lab}}_{i}(\tau^k)\otimes\hat{E}^{k'\,{\rm lab}}_{j}(\tau^{k'}).
\end{equation}
$\hat{E}_{i}^{k\,\rm lab}(\tau)$ are the Kraus operators governing
the time evolution of a one-particle system, and $i,j=0,\hdots,5$
(see Appendix~A), and $\hat{E}_0^{k\,\rm lab}(\tau)=\hat{E}_0$. The
normalization condition $\sum_{k\in Q,k'\in
Q'}\sum_{i,j}(\hat{E}_{ij}^{kk'}(t))^{\dag}\hat{E}_{ij}^{kk'}(t)=
\id\otimes\id$
is fulfilled, which follows from corresponding normalization
conditions for one-particle Kraus operators (\ref{norm}). 
It is clear that for \emph{distinguishabl}e particles we have
$Q=\{p\}$ and $Q'=\{q\}$. However in the case of \emph{identical} 
particles one does not know which
of them carries which four-momentum or flavour, thus the space of the
whole system must be symmetrized, and $Q=Q'=\{p,q\}$. Let us
introduce the permutation operator $\mathcal{P}$ such that
$\mathcal{P}(\ket{s,k}\otimes\ket{s',k'})
=\ket{s',k'}\otimes\ket{s,k}$; then
\begin{equation}
\mathcal{P}\hat{\varrho}^{\rm
lab}_{\mathcal{A}\mathcal{B}}(t)\mathcal{P}=\hat{\varrho}^{\rm
lab}_{\mathcal{A}\mathcal{B}}(t). \label{symetryczna mg}
\end{equation}
Moreover any observable $\hat{O}$ must also preserve symmetrization,
i.e.:
\begin{equation}
\label{obserwable} \mathcal{P}\;\hat{O}\;\mathcal{P}=\hat{O}.
\end{equation}
One can easily verify that if $\hat{\varrho}^{\rm
lab}_{\mathcal{A}\mathcal{B}}(0)$ fulfils the condition
(\ref{symetryczna mg}), than also $\hat{\varrho}^{\rm
lab}_{\mathcal{A}\mathcal{B}}(t)$ obtained from (\ref{krausABa})
fulfils it, so the time evolution of identical particles is governed
by the same evolution law (\ref{krausABa}) as the time evolution of
the distinguishable ones.

In EPR-type experiments it is of great importance that the
measurements performed by distant observers, say Alice and Bob, must
be local. The locality means that the observables are restricted to
some specific region usually interpreted as the region of detector
(see e.g.~\cite{caban03a}). It can be achieved experimentally by
assuming, for example, that Alice's detector can register only
particles with four-momentum $p$ and Bob's only those with
four-momentum $q$ (directions of momenta $\vec{p}$ and $\vec{q}$ must
be sufficiently different)\footnote{Actually the sharp momentum states cannot
  be achieved because of the uncertainty principle and finite volume
  of detector.}.  Thus, only the particle carrying
four-momentum $p$ and $q$, respectively can reach Alice's and Bob's
detectors.

The space of states of a two-particle 
system is the tensor product
$\left(\mathcal{H}_{\mathcal{A}}\oplus\mathcal{H}_{0}\right)\otimes
\left(\mathcal{H}_{\mathcal{B}}\oplus\mathcal{H}_{0}\right)$ (see
(\ref{hilbert})). In the case of distinguishable particles
$\mathcal{H}_{\mathcal{A}}$ is spanned by $\{\ket{s,p}\}$ and
$\mathcal{H}_{\mathcal{B}}$ by $\{\ket{s',q}\}$, so Alice's detector 
registers only the particles with states from subspace
$\mathcal{H}_{\mathcal{A}}\oplus\mathcal{H}_{0}$ and Bob's only 
those from $\mathcal{H}_{\mathcal{B}}\oplus\mathcal{H}_{0}$.
On the other hand, in the case of indistinguishable particles
$\mathcal{H}_{\mathcal{A}}$ and $\mathcal{H}_{\mathcal{B}}$ must be
identical and both spanned by the same set of vectors $\{\ket{s,p},
\ket{s',q}\}$. The physical Hilbert space is symmetric subspace of 
$\mathcal{H}$, i.e. $\tfrac{1}{2}(\id+\mathcal{P})\mathcal{H}$,
therefore one cannot associate specific one-particle Hilbert space
with a given observer.

\section{Quantum correlations in the neutral kaon system}
\label{correlations} Let us consider two neutral kaons in a given
initial state $\hat{\varrho}^{\rm lab}_{\mathcal{A}\mathcal{B}}(0)$,
and two distant observers, Alice and Bob, in the same laboratory
frame. Alice can measure the flavour of kaons only with four-momentum
$p$ and Bob only those with four-momentum $q$, therefore their
observables commute, i.e.:
\begin{equation}
\label{komutator} [\hat{A}^p,\hat{B}^q]=0.
\end{equation}
Hereafter, we omit the superscript `{\rm lab}', because in the sequel we
consider only density operators as seen from the laboratory frame.

\subsection{Distinguishable case}
\label{dist} In this subsection we treat the kaons as
distinguishable particles, as it is usually done to simplify
calculations.  
The Hilbert space of the first kaon is spanned by set of
orthonormal vectors
$\left\{\ket{K^0,p},\ket{\bar{K}^0,p},\ket{0,0}\right\}$ and the
Hilbert space of the second one by
$\left\{\ket{K^0,q},\ket{\bar{K}^0,q},\ket{0,0}\right\}$.

Suppose that at time $t_A$ Alice measures an observable
$\hat{A}^p=A\nolinebreak\otimes\nolinebreak\id$ with the spectral
decomposition $\sum_{a}a \left(\Pi_a\otimes \id\right)$. The density
operator just before the measurement is
\begin{equation}
\label{ewolucja z czasem wlasnym}
\hat{\varrho}_{\mathcal{A}\mathcal{B}}(t_A)=\sum_{ij}\hat{E}_{ij}^{pq}(t_A)
\hat{\varrho}_{\mathcal{A}\mathcal{B}}(0)(\hat{E}_{ij}^{pq}(t_A))^{\dag},
\end{equation}
(cf.~(\ref{krausABa})), and when the outcome of the measurement is
$a$, the state reduces to
\begin{equation}
\label{rho a}
\hat{\varrho}_{\mathcal{A}\mathcal{B}}^a(t_A)=\frac{(\Pi_a\otimes
\id) \hat{\varrho}_{\mathcal{A}\mathcal{B}}(t_A)(\Pi_a\otimes
\id)}{p_a(t_A)},
\end{equation}
where $p_a(t_A)$ is the probability of measuring $a$ at time $t_A$.
Next, at time $t_B$, Bob performs the measurement of
$\hat{B}^q=\id\otimes B$ with spectral decomposition $\sum_b b
\left(\id\otimes\Pi_b\right)$. Just before his measurement the state
is
\begin{equation}
\label{ewolucja do tb}
\hat{\varrho}_{\mathcal{A}\mathcal{B}}^a(t_B)=\sum_{ij}\hat{E}_{ij}^{pq}
(t_B-t_A)\hat{\varrho}_{\mathcal{A}\mathcal{B}}^a(t_A)
(\hat{E}^{pq}_{ij}(t_B-t_A))^{\dag}.
\end{equation}
The conditional probability that Bob's outcome is $b$ provided that
Alice's was $a$ is
\begin{equation}
\label{prob warunkowe} p_{b|a}(t_B)=\tr\left\{\left(\id\otimes\Pi_b
\right)\hat{\varrho}_{\mathcal{A}\mathcal{B}}^a(t_B)\left(
\id\otimes \Pi_b\right)\right\}.
\end{equation}
By means of (\ref{ewolucja z czasem wlasnym})--(\ref{prob
warunkowe}), the joint probability
\begin{equation}
p_{ab}(t_A,t_B)=p_a(t_A) p_{b|a}(t_B), \label{prob calkowite}
\end{equation}
that Alice's and Bob's outcomes are $a$ and $b$, respectively 
is given by the formula
\begin{equation}
\label{prob ostateczne}
p_{ab}(t_A,t_B)=\tr\bigl\{\hat{\varrho}_{\mathcal{A}\mathcal{B}}(0)
\bigl[\sum_{i}(\hat{E}_{i}^{p}
\left(\tau_A^p\right))^{\dag}\Pi_a
\hat{E}_{i}^{p}\left(\tau_A^p\right) \otimes
\sum_{j}(\hat{E}_{j}^{q}\left(\tau_B^q\right))^{\dag}\Pi_b
\hat{E}_{j}^{q}\left(\tau_B^q\right) \bigr]\bigr\}.
\end{equation}
Then the correlation function between the outcomes
\begin{equation}
\label{kor} C_{A^pB^q}(t_A,t_B)=\sum_{ab}ab \;p_{ab}(t_A,t_B),
\end{equation}
takes the form
\begin{align}
\label{korelacje}
C_{A^pB^q}(t_A,t_B)=\tr\left\{\hat{\varrho}_{\mathcal{A}\mathcal{B}}(0)
\left[A(\tau_A^p)\otimes
B(\tau_B^q)\right]\right\},
\end{align}
where
$A(\tau_A^p)=\sum_{i}(\hat{E}_{i}^{p}\left(\tau_A^p\right))^{\dag}A
\hat{E}_{i}^{p}\left(\tau_A^p\right)$, and
$B(\tau_B^q)=\sum_{j}(\hat{E}_{j}^{q}\left(\tau_B^q\right))^{\dag}B
\hat{E}_{j}^{q}\left(\tau_B^q\right).$

Now  we calculate explicitly a few correlation functions and
probabilities in the system of neutral kaons in the pure entangled 
state $J^{PC}=1^{--}$ (produced through the reaction
$e^{+} e^{-} \to \phi(1020) \to K^{0} \bar{K}^{0}$) 
\begin{equation}
\label{stan}
\ket{\psi}=\frac{1}{\sqrt{2}}\left(\ket{K^0,p}
\otimes\ket{\bar{K}^0,q}-\ket{\bar{K}^0,p}\otimes\ket{K^0,q}\right)
\end{equation}
to show how our model works. We do not neglect CP violation and
decoherence, despite the fact that CPLEAR experiment 
at CERN (where $K^{0} \bar{K}^{0}$ pairs are produced in the 
$p \bar{p}$ collider) is not sensitive to CP violating effects. 
Let us begin with the strangeness
correlations. The strangeness operator takes
the form  
\begin{equation}
\label{dziwnoscq}
S^k=\ket{K^0,k}\bra{K^0,k}-\ket{\bar{K}^0,k}\bra{\bar{K}^0,k},
\end{equation}
where $k=p,q$, and $\hat{A}^p = S^p \otimes \id$,    
$\hat{B}^q = \id \otimes S^q$. It is easy to show that the
corresponding correlation function has the following form
\begin{multline}
\label{korelacjeSS}
C_{S^pS^q}(t_A,t_B)=- \frac{1}{1-\delta_L^2}
\bigl[\e^{-(\Gamma+\lambda)(\tau_A^p+\tau_B^q)}\cos(\Delta
m \Delta \tau)\\
-\tfrac{1}{2}\delta_L^2\left(\e^{-\Gamma_S
\tau_A^p-\Gamma_L \tau_B^q}+\e^{-\Gamma_L \tau_A^p-\Gamma_S
\tau_B^q}\right)\bigr],
\end{multline}
where $\tau_A^p$ and $\tau_B^q$ stand for the proper times, 
$\Delta\tau=\tau_B^q-\tau_A^p$, 
$\delta_L = 2{\Re}(\epsilon)/(1+|\epsilon|^2)$ 
($\epsilon$ is a small complex
$CP$-violation parameter), $\Gamma=\half(\Gamma_S+\Gamma_L)$
($\Gamma_S$ and $\Gamma_L$ are the decay widths of short and long
living states of neutral kaon, respectively),
$\Delta m=m_L-m_S$ 
($m_S$ and $m_L$ are masses of short and long
living states of neutral kaon, respectively)
and $\lambda$ is a decoherence parameter, 
representing interaction between one-particle
system and the environment. 
On the other hand, the strangeness operators defined
above have three different eigenvalues $±1$ and $0$, but the
observables considered in Bell-CHSH inequalities 
have only two different eigenvalues $± 1$. Such a dichotomic
observable is, for example, an observable answering the question
whether one registers a kaon (anti-kaon) (then the result of the
measurement is $+1$), or not (then the result is $-1$). Such a case
was investigated in \cite{bertlmann06}. Denoting this observable as
$D_+$ ($D_-$), in one-particle case we have
\begin{equation}
\label{d+-}
D_{±}^k=±\ket{K^0,k}\bra{K^0,k}\mp\ket{\bar{K}^0,k}
\bra{\bar{K}^0,k}-\ket{0,0}\bra{0,0}.
\end{equation}
It is easy to show that in the case when
$\hat{A}^p=D^p_+\otimes\id$ and $\hat{B}^q=\id\otimes D^q_+$, we get
\begin{multline}
\label{korelacjeDD}
C_{D_+^pD_+^q}(t_A,t_B)=1-\frac{1+\delta_L}{1-\delta_L}
\bigl[\e^{-(\Gamma+\lambda)(\tau_A^p+\tau_B^q)}\cos(\Delta
m \Delta \tau)\\
-\tfrac{1}{2}\left(\e^{-\Gamma_S \tau_A^p-\Gamma_L
\tau_B^q}+\e^{-\Gamma_L \tau_A^p-\Gamma_S
\tau_B^q}\right)\bigr] \\
-\frac{1}{2(1-\delta_L)}\left(\e^{-\Gamma_S
\tau_A^p}+\e^{-\Gamma_S \tau_B^q}+\e^{-\Gamma_L \tau_A^p}+\e^{-\Gamma_L
\tau_B^q}\right)\\
+\frac{\delta_L}{1-\delta_L}\left(\e^{-(\Gamma+\lambda)\tau_A^p}
\cos(\Delta m \tau_A^p)+\e^{-(\Gamma+\lambda)\tau_B^q}\cos(\Delta
m \tau_B^q)\right),
\end{multline}
and when $\hat{A}^p=D^p_+\otimes\id$ and $\hat{B}^q=\id\otimes
D^q_-$, we have
\begin{multline}
\label{korelacjed+d-}
C_{D_+^pD_-^q}(t_A,t_B)=1+\e^{-(\Gamma+\lambda)(\tau_A^p+\tau_B^q)}
\cos(\Delta m \Delta \tau) \\
- \frac{1}{2(1-\delta_L)}\left(\e^{-\tau_A^p
\Gamma_S}+\e^{-\tau_A^p \Gamma_L}-2\delta_L \e^{-\tau_A^p
(\Gamma+\lambda)}\cos(\Delta m \tau_A^p)\right)\\
-
\frac{1}{2(1+\delta_L)}\left(\e^{-\tau_B^q \Gamma_S}+\e^{-\tau_B^q
\Gamma_L}+2\delta_L \e^{-\tau_B^q (\Gamma+\lambda)}\cos(\Delta m
\tau_B^q)\right)\\
+ \tfrac{1}{2}\left(\e^{-\tau_A^p
\Gamma_S-\tau_B^q\Gamma_L}+\e^{-\tau_A^p \Gamma_L-\tau_B^q
\Gamma_S}\right).
\end{multline}
Of course, we could find the above quantum correlation
functions directly from the definition (\ref{kor}) finding appropriate
probabilities. For example, one can easily find the stangeness
correlation function $C_{S^pS^q}(t_A,t_B)$ knowing that\\
\begin{subequations}
\label{prawdopodobienstwo}
\noindent(i) the probability that Alice's detector registers 
$K^0$ at $t_A$ and Bob's $K^0$ at $t_B$ is
\begin{multline}
\label{prawdopodobienstwoKK}
p_{K^0,K^0}(t_A,t_B)=\frac{1}{8}\frac{1+\delta_L}{1-\delta_L}
\bigl[\e^{-\Gamma_S \tau_A^p-\Gamma_L \tau_B^q}
+\e^{-\Gamma_S \tau_B^q-\Gamma_L \tau_A^p}\\
-2\e^{-(\Gamma+\lambda)(\tau_A^p+\tau_B^q)}\cos(\Delta m \Delta \tau)\bigr]\,,
\end{multline}
(ii) the probability that Alice's detector registers $\bar{K}^0$ at
$t_A$ and Bob's $\bar{K}^0$ at $t_B$ is
\begin{multline}
\label{prawdopodobienstwoAntyKAntyK}
p_{\bar{K}^0,\bar{K}^0}(t_A,t_B)=\frac{1}{8}\frac{1-\delta_L}{1+\delta_L}
\bigl[\e^{-\Gamma_S \tau_A^p-\Gamma_L \tau_B^q}+
\e^{-\Gamma_S \tau_B^q-\Gamma_L \tau_A^p}\\
-2\e^{-(\Gamma+\lambda)(\tau_A^p+\tau_B^q)}\cos(\Delta m \Delta \tau)\bigr]\,,
\end{multline}
(iii) the probabilities that Alice's detector registers 
$\bar{K}^0$ at $t_A$ and Bob's $K^0$ at $t_B$, and that 
Alice's detector registers $K^0$ at $t_A$ and
Bob's $\bar{K}^0$ at $t_B$ are 
\begin{equation}
\begin{split}
\label{prawdopodobienstwoKAntyK}
p_{\bar{K}^0,K^0}(t_A,t_B)&=p_{K^0,\bar{K}^0}(t_A,t_B)\\ 
&=\frac{1}{8}\left[\e^{-\Gamma_S
\tau_A^p-\Gamma_L \tau_B^q}+ \e^{-\Gamma_S \tau_B^q-\Gamma_L
\tau_A^p}+2\e^{-(\Gamma+\lambda)(\tau_A^p+\tau_B^q)}\cos(\Delta m
\Delta \tau)\right] \,.
\end{split}
\end{equation}
\end{subequations}
In \cite{bertlmann06} the correlation function
$C_{D_+^pD_+^q}(t_A,t_B)$ and probabilities  $p_{K^0,K^0}(t_A,t_B)$,
$p_{\bar{K}^0,\bar{K}^0}(t_A,t_B)$ and $p_{\bar{K}^0,K^0}(t_A,t_B)$ 
were found under condition $\delta_L=\lambda=0$,  
i.e.~without CP-violation and decoherence. 
Of course, when we put $\delta_L=\lambda=0$ in (\ref{korelacjeDD}) 
and (\ref{prawdopodobienstwo}) we arrive at Bertlmann's results, 
if we only take into account differences in conventions.

\subsection{Indistinguishable case}
Now let us  
consider the same situation as in the previous subsection but with
more realistic assumption that kaons are indistinguishable particles.
Suppose that Alice measures $\hat{A}^p$ at time $t_A$ and Bob
measures $\hat{B}^q$ at time $t_B$. Both of the observables must
fulfil (\ref{obserwable}). After analogous calculations as in the
previous subsection, using (\ref{komutator}), we get 
\begin{equation}
\label{kor_identyczne}
C_{\hat{A}^p\hat{B}^q}(t_A,t_B)=
\tr\left[\hat{\varrho}_{\mathcal{A}\mathcal{B}}(0)\hat{A}^p(\tau_A^p)
\hat{B}^q(\tau_B^q)\right],
\end{equation}
where 
\[\hat{A}^p(\tau_A^p)=\sum_{ij}(\hat{E}_{ij}^{pq}(t_A))^{\dag}
\hat{A}^p\hat{E}_{ij}^{pq}(t_A)
\quad\text{and}\quad
\hat{B}^q(\tau_B^q)=\sum_{ij}(\hat{E}^{pq}_{ij}(t_B))^{\dag}
\hat{B}^q\hat{E}_{ij}^{pq}(t_B)\,.\]

Now let us calculate the same correlation functions as before. 
First, we have to note that in the case of indistinguishable 
particles the initial state has different form   
\begin{multline}
\label{psi}
\ket{\psi}=\frac{1}{2}\left(\ket{K^0,p}\otimes\ket{\bar{K}^0,q}+
\ket{\bar{K}^0,q}\otimes\ket{K^0,p}
\right.\\-
\left.\ket{\bar{K}^0,p}\otimes\ket{K^0,q}-\ket{K^0,q}\otimes
\ket{\bar{K}^0,p}\right).
\end{multline}
Second, we have to construct observables that answer the same
questions as the observables used in the case of distinguishable
kaons. The strangeness operators $\hat{S}^k$ take the form
\begin{equation}
\label{dziwnosc} \hat{S}^k=S^k\otimes \id+\id\otimes S^k,
\end{equation}
where $k$ takes the value $p$ or $q$, and $S^k$ was defined in the
previous subsection. Observables $\hat{D}^k_{±}$ cannot be
constructed in analogy to (\ref{dziwnosc}) by means of (\ref{d+-})
and symmetrization\footnote{When we calculate the spectral
decomposition of, say, $D^k_+\otimes \id+\id\otimes D^k_+$, where
$D^k_+=2\ket{K^0,k}\bra{K^0,k}-\id$, we find out that it has
eigenvalues equal $±2$ and $0$, so it does not answer the
question, whether the particle is kaon carrying momentum $k$, or
not.}. Now, the form of $\hat{D}^k_{+}$ is
\begin{multline}
\label{d+k} \hat{D}^k_+=2\left(\ket{K^0,k}\bra{K^0,k}\otimes
\id+\id\otimes \ket{K^0,k}\bra{K^0,k}\right)\\-\id\otimes \id
-\ket{K^0,k}\bra{K^0,k}\otimes \ket{K^0,k}\bra{K^0,k}.
\end{multline}
It yields $+2$ when both of the particles are kaons with
four-momentum $k$, $+1$ when one of the particles is a kaon with
four-momentum $k$, and $-1$, when there is no such kaon. It is not
dichotomic, but it is not a problem because in state (\ref{psi}) the
probability of measuring two kaons with the same four-momentum
equals zero. $\hat{D}^k_-$ takes the analogous form to $\hat{D}^k_{+}$
\begin{multline}
\label{d-k}
\hat{D}^k_-=2\left(\ket{\bar{K}^0,k}\bra{\bar{K}^0,k}\otimes
\id+\id\otimes
\ket{\bar{K}^0,k}\bra{\bar{K}^0,k}\right)\\-\id\otimes \id
-\ket{\bar{K}^0,k}\bra{\bar{K}^0,k}\otimes
\ket{\bar{K}^0,k}\bra{\bar{K}^0,k}.
\end{multline}
It is easy to check that the correlation functions
$C_{\hat{S}^p\hat{S}^q}(t_A,t_B)$,
$C_{\hat{D}_+^p\hat{D}_+^q}(t_A,t_B)$ and
$C_{\hat{D}_+^p\hat{D}_-^q}(t_A,t_B)$ are exactly the same as in
the distinguishable particles case (\ref{korelacjeSS}),
(\ref{korelacjeDD}) and (\ref{korelacjed+d-}).
It is quite remarkable that for the observables considered above,
it does not matter whether we treat kaons as indistinguishable particles
or not, at least in the singlet state. Of course, it is easier to
carry out all calculations on the assumption that kaons are
distinguishable particles, as it is usually done.

\section{Conclusions}
We have shown that having the Kraus representation of time evolution
of an unstable particle in its rest frame it is possible to find the
evolution of the particle in an arbitrary reference frame. Moreover, 
we have also shown that taking into account the form of Kraus
operators one can extend the time evolution of the system 
to the case in which the state of the particle is the superposition 
or mixture of a few different momentum eigenstates. Next, we have
found the time evolution of two non-interacting unstable particles,
distinguishable as well as identical ones. Finally, we have applied 
these results in calculation of some quantum correlation functions for 
$K^0 \bar{K}^0$ system in the singlet state assuming $CP$-violation
and decoherence. 
And it turned out that the results are exactly the same either 
we treat kaons as distinguishable particles or identical ones. 
Therefore, one can neglect the fact that kaons are identical
particles and treat them as distinguishable ones, at least in the cases 
considered in this paper. 
It is still an open question whether the statistics of the particles
plays no role in the general case or not.

We would like to point out that all results presented in this
paper will be valid also for B-mesons after appropriate change of
notation, because kaons and B-mesons evolve according to the same
scheme.

\begin{ack}
  We would like to acknowledge fruitfull discussions with R.~Alicki
  and A.~Kossakowski.  This work is supported by the Polish Ministry
  of Scientific Research and Information Technology under the grant
  No.~PBZ/MIN/008/P03/2003 and by the University of Lodz.
\end{ack}

\appendix
\section{Time evolution of $K^0$}
\label{AppA} Let us assume that the space of states of the system is
spanned by the orthogonal vectors: vacuum $\ket{0,0}$ and
$\ket{s,k}$, where $s$ denotes inner degrees of freedom and
four-momentum $k$ belongs to the finite set $Q$ of admissible
four-momenta. The Kraus operators $\hat{E}_i^k(t)=
\sum_{s,s'}E^{ss'}_i(\tau^k)\ket{s,k}\bra{s',k}$, describing evolution
(\ref{ewolucja_wiele_pedow}) of the kaon which state can be a
superposition (or mixture) of different momentum
eigenstates, have the following form 
\allowdisplaybreaks
\begin{subequations}
  \label{eq:kaon-Krauss}
  \begin{align}
    \hat{E}_0^k(\tau^k) &=\hat{E}_0=\ket{0,0}\bra{0,0}\,,\\
    \hat{E}_1^k(\tau^k) &= \frac{1}{2}\left(\e^{-\tau^k(\lambda+2im_S+\Gamma_S)/2}
    +\e^{-\tau^k(\lambda+2im_L+\Gamma_L)/2}\right) \notag\\&\qquad×
    \left(\ket{K^0,k}
    \bra{K^0,k}+\ket{\bar{K}^0,k}\bra{\bar{K}^0,k}\right)\notag\\&+
    \frac{1}{2}\left(\e^{-\ \tau^k(\lambda+2im_S+\Gamma_S)/2}
    -\e^{-\tau^k(\lambda+2im_L+\Gamma_L)/2}\right)\notag\\&\qquad×
    \left(\frac{1+\epsilon}{1-\epsilon}
    \ket{K^0,k}
    \bra{\bar{K}^0,k}+\frac{1-\epsilon}{1+\epsilon}\ket{\bar{K}^0,k}
    \bra{K^0,k}\right)\,,\\
    \label{eq:KraussE1}
    \hat{E}_2^k(\tau^k) &= \sqrt{\frac{1+|\epsilon^2|}{2}}
    \left(1 - \e^{-\tau^k \Gamma_S} - \delta_L^2 \frac{\left|1
          - \e^{-\tau^k (\Gamma + \lambda - i \Delta m)}
    \right|^2}{1 - \e^{-\tau^k \Gamma_L}}\right)^{\frac{1}{2}}\notag\\&\qquad×
    \left(\frac{1}{1+\epsilon}\ket{0,0} \bra{K^0,k} 
      +\frac{1}{1-\epsilon}\ket{0,0} \bra{\bar{K}^0,k}\right)\,,\\
    \hat{E}_3^k(\tau^k) &= 
    \sqrt{\frac{1+\left|\epsilon\right|^2}{2 (1 - \e^{-\tau^k\Gamma_L})}}
    \left(
    \frac{1-\e^{-\tau^k \Gamma_L}+
    \delta_L-\e^{-\tau^k(\lambda-i\Delta
      m+\Gamma)}\delta_L}{1+\epsilon}
    \ket{0,0} \bra{K^0,k}\right.
    \notag\\* &\quad\left.-\frac{1-\e^{-\tau^k
    \Gamma_L}-
    \delta_L+\e^{-\tau^k(\lambda-i\Delta
      m+\Gamma)}\delta_L}{1-\epsilon}
    \ket{0,0} \bra{\bar{K}^0,k}\right) \,,\\
      \hat{E}_4^k(\tau^k) &=
      \frac{1}{2}\e^{-\tau^k\Gamma_S/2}
    \sqrt{1-\e^{-\tau^k\lambda}}\left(\ket{K^0,k}\bra{K^0,k}+
      \ket{\bar{K}^0,k}\bra{\bar{K}^0,k}
    \right.\notag\\* &\quad\left.
    +\frac{1+\epsilon}{1-\epsilon}\ket{K^0,k}\bra{\bar{K}^0,k} 
      +\frac{1-\epsilon}{1+\epsilon}\ket{\bar{K}^0,k}\bra{K^0,k}
      \right)\,,\\
      \hat{E}_5^k(\tau^k) &=\frac{1}{2}\e^{-\tau^k\Gamma_L/2}
    \sqrt{1-\e^{-\tau^k\lambda}}\left(\ket{K^0,k}\bra{K^0,k}+
      \ket{\bar{K}^0,k}\bra{\bar{K}^0,k}
    \right.\notag\\* &\quad\left.
      -\frac{1+\epsilon}{1-\epsilon}\ket{K^0,k}\bra{\bar{K}^0,k} 
      -\frac{1-\epsilon}{1+\epsilon}\ket{\bar{K}^0,k}\bra{K^0,k}
      \right)\,,
    \end{align}
\end{subequations}
where $\tau^k=t/\gamma^k$, $\epsilon$ is a small complex
$CP$-violation parameter, 
$\delta_L = 2 {\Re}(\epsilon)/(1+|\epsilon|^2)$, 
$\Gamma_S$ and $\Gamma_L$
are the decay widths of $K^0_S$ and $K^0_L$ (short and long living
states of neutral kaon), respectively,
$\Gamma=\half(\Gamma_S+\Gamma_L)$, 
$m_S$ and $m_L$ are masses of $K^0_S$ and $K^0_L$, respectively, 
$\Delta m=m_L-m_S$, and $\lambda$ is a
decoherence parameter, representing interaction between one-particle
system and the environment. 
In comparison to \cite{caban05} we use different, more
convenient set of Kraus operators, leading to the same evolution.
It is easy to check that the normalization condition
\begin{equation}
\label{norm}
  \hat{E}_{0}^\dag
 \hat{E}_{0}
  + \sum_{k\in Q}\sum_{i=1}^{5}(\hat{E}^{k\,{\rm lab}}_{i}(\tau^k))^\dag
 \hat{E}^{k\, {\rm lab}}_i(\tau^k)=\id
 \end{equation}
holds. In the case of indistinguishable particles the Kraus
operators are exactly the same, but we additionally must sum over
all admissible $k$.

\bibliographystyle{elsart-num}

\end{document}